\documentclass[showkeys,showpacs]{revtex4}
\usepackage{graphicx}
\usepackage{amsmath}
\usepackage{amssymb}
\usepackage{color}

\newtheorem{theorem}{Theorem}[section]
\newtheorem{definition}{Definition}[section]

\begin{document}

\title{Quantum wormhole as a Ricci flow}

\author{Vladimir Dzhunushaliev
\footnote{Senior Associate of the Abdus Salam ICTP}}
\email{vdzhunus@krsu.edu.kg}
\affiliation{Department of Physics and Microelectronic
Engineering, Kyrgyz-Russian Slavic University, Bishkek, Kievskaya Str.
44, 720021, Kyrgyz Republic}

\begin{abstract}
The idea is considered that a quantum wormhole in a spacetime foam can be described as a Ricci flow. In this interpretation the Ricci flow is a statistical system and every metric in the Ricci flow is a microscopical state. The probability density of the microscopical state is connected with a Perelman's functional of a rescaled Ricci flow. 
\end{abstract}

\pacs{04.50.+h,02.90.+p,04.90.+e\\
2000 MSC: 53C44, 53C21, 83D05, 83E99}
\keywords{Ricci flows, quantum wormhole}

\maketitle

\section{Introduction}

\begin{figure}[h]
  \includegraphics[width=7cm,height=6cm]{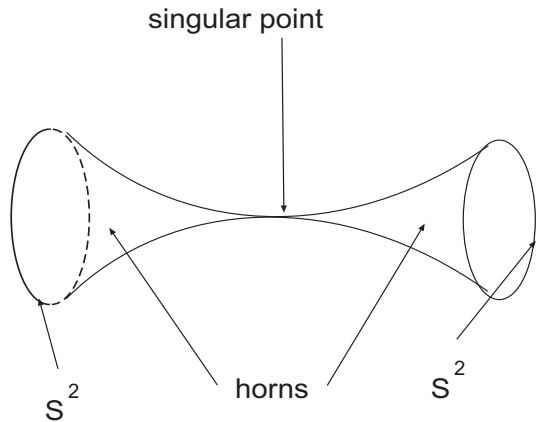}
  \caption{The horns and singular point in a Ricci flow.}
\label{horns}
\end{figure}

Ricci flows are the tool for the investigation of the topology of manifolds. Generally Ricci flow creates on a manifold a singularity (or singularities) for a finite parameter $\lambda_0$. Fig.~\ref{horns} gives a schematic picture of the partially singular metric $g(\lambda_0)$ on the manifold $\mathcal M$. The metric $g(\lambda_0)$ is smooth on a maximal domain 
$\Omega \subset \mathcal M$, where the curvature is locally bounded but is singular, i.e. ill-defined, on the complement where the curvature blows-up as $\lambda \rightarrow \lambda_0$.

\par 
If $\Omega \neq \emptyset$, then the main point is that small neighborhoods of the boundary $\partial \Omega$ consist of horns. A horn is a metric on $S^2 \times [0, \delta]$ where the $S^2$ factor is approximately round of radius $\rho(r)$, with $\rho(r)$ small and 
$\rho(r)/r \rightarrow 0$ as $r \rightarrow 0$. Fig.~\ref{horns} represents a partially singular metric on the smooth manifold $S^2 \times I$, consisting of a pair horns joined by a degenerate metric. 
\par 
This is exactly the same what physicists are talking about a quantum wormhole in a spacetime foam. It allows us to think that Ricci flows are the mathematical tool for the description of the  wormhole in the spacetime foam. Here we offer the idea that for every $\lambda$ the 3D space-like metric $g(\lambda)$ is realised with some probability $\rho(\lambda)$ where the parameter $\lambda$ describes the evolution of the metric $g$ under the Ricci flow. Usually such probability is connected with path integral. We offer the idea that this probability is connected with a Perelman's functional $\mathcal W$ on a rescaled Ricci flow as $\rho \propto \frac{d \mathcal W}{d \lambda}$ in the consequence of the property $\frac{d \mathcal W}{d \lambda} \geq 0$. Then the Ricci flow is a statistical system where every metric $g(\lambda)$ is a microscopical state. 
\par 
The notion of a spacetime foam was introduced by Wheeler \cite{wheel1} for the description of the possible complex structure of spacetime on the Planck scale 
($l_{Pl} \approx 10^{-33}cm$). The exact mathematical description of this phenomenon is very difficult and even though there is a doubt: does the Feynman path integral in the gravity contain a topology change of the spacetime ? This question spring up as (according to the Morse theory) the singular points must arise by topology changes. In such points the time arrow is undefined that leads in difficulties at the definition of the Lorentzian metric, curvature tensor
and so on. 
\par 
The Ricci flows were introduced by Hamilton \cite{Hamilton} over 25 years ago. It plays an important role in the proof of the Poincare conjecture \cite{Perelman}. In Ref.~\cite{Husain:2008rg} the evolution of wormhole geometries under Ricci flow is studied. Depending on value of initial data parameters, wormhole throats either pinch off or evolve to a monotonically growing state. The connection between Ricci flow and quantum mecanins is considered in Ref's~\cite{Isidro:2008ik} \cite{Isidro:2008nh}. The physical application of Ricci flow can be found in Ref's~\cite{Headrick:2006ti} - \cite{Woolgar:2007vz} as well. Very simple introduction to Ricci flows for non-specialists can be found in Ref.~\cite{anderson}. 
\par 
Quantum gravity remains a theorists playground, an arena for theoretical experiments which may or may not stand the test of time. The two leading candidates for a quantum theory of gravity  today are string theory and loop quantum gravity. One can find a thorough introduction to string theory in textbook \cite{Polchinski}, and a review of loop quantum gravity in Ref. \cite{Rovelli}.  Despite some 70 years of active research, no one has yet formulated a consistent and complete quantum theory of gravity. The failure to quantize gravity rests in part on technical difficulties. General relativity is a complicated and highly nonlinear theory. But the real problems are almost certainly  deeper: quantum gravity requires a quantization of spacetime itself, and at a fundamental level we do not know what that means.
\par 
Our point of view is that in quantum gravity it is necessary to separate two problems: the first one is the quantization of very strongly self-interacting gravitational field (metric), the second one is the problem of topology change in quantum gravity. We think that it is two different problems. For the first problem we need a non-perturbative quantization technique (in some more weakly sense this problem one can find in quantum chromodynamics -- so called confinement problem). In this paper we propose an approach for the second problem: we conjecture that a quantum wormhole in a spacetime foam may be described on the Ricci flow language. The mathematicians use this language for the description of topology change. Our  approach is that the quantum wormhole can be described as a statistical system and corresponding probability is connected with Perelman functional. 

\section{Ricci flows}

In this section we follow to Ref.~\cite{topping}. Ricci flow is a means of processing the metric $g_{ab}$ by allowing it to evolve under 
\begin{equation}
	\frac{\partial g_{ab}(x^c, \lambda)}{\partial \lambda} = -2 R_{ab}(x^c, \lambda)
\label{ric-10}
\end{equation}
where $R_{ab}$ is the Ricci curvature; $\lambda$ is a parameter; $a,b = 1,2,3$; $x^c$ is the coordinate on a manifold $\mathcal M$. The Ricci flow describes the evolution of the metric $g_{ab}$ in during of the parameter $\lambda$. In Ref.~\cite{topping} such reply for the question ``Ricci flow: what is it, and from where did it come'' is given: ``$\ldots$ the flow can be used to deform $g_{ab}$ into a metric distinguished by its curvature. For example, if $\mathcal M$ is two-dimensional, the Ricci flow deforms a metric conformally to one of constant curvature, and thus gives a proof of the two-dimensional uniformisation theorem. More generally, the topology of $\mathcal M$ may preclude the existence of such distinguished metrics, and the Ricci flow can then be expected to develop a singularity in finite time\footnote{In our notations the time is the parameter $\lambda$.}. Nevertheless, the behavior of the flow may still serve to tell us much about the topology of the underlying manifold. The strategy of the investigation is to stop a flow, and then carefully perform ``surgery'' on the evolved manifold, exciting any singular regions before continuing the flow. 
\par 
Provided we understand the structure of finite time singularities sufficiently well, we may hope to keep track of the change in topology of the manifold under surgery, and reconstruct the topology of the original manifold from a limiting flow, together with the singular regions removed.''
\par 
Let us introduce the functional
\begin{equation}
	\mathcal W(g_{ab}, f, \tau) := 
	\int \left[
		\tau \left( R + \left| f \right|^2 \right) + f - n
	\right] \, u \; dV
\label{ric-20}
\end{equation}
where $f: \mathcal M \rightarrow \mathbb R$ is a smooth function; $R$ is the Ricci scalar; $\tau > 0$ is a scale parameter; $n = \mathrm{dim} \mathcal M$, and $u$ is defined by 
\begin{equation}
	u:= (4 \pi \tau)^{-n/2} e^{-f}. 
\label{ric-30}
\end{equation}
\begin{theorem}
If $\mathcal M$ is closed, and $g_{ab}, f$ and $\tau$ evolve according to 
\begin{eqnarray}
	\frac{\partial g_{ab}}{\partial \lambda} &=& -2 R_{ab},
\label{ric-40}\\
	\frac{d \tau}{d \lambda} &=& -1,
\label{ric-50}\\
	\frac{\partial f}{\partial \lambda} &=& 
	- \Delta f + \left| \nabla f \right|^2 - R + 
	\frac{n}{2 \tau}
\label{ric-60}
\end{eqnarray}
then the functional $\mathcal W$ increases according to 
\begin{equation}
	\frac{d}{d \lambda} \mathcal W(g_{ab}, f, \tau) = 
	2 \tau \int \left|
		R_{ab} + \frac{\partial^2 f}{\partial x^a \partial x^b} - 
		\frac{g_{ab}}{2 \tau}
	\right|^2 u dV \geq 0
\label{ric-70}
\end{equation}
where $R_{ab}$ is the Ricci tensor. 
\end{theorem}
Under evolution \eqref{ric-40}-\eqref{ric-60} 
\begin{equation}
	\frac{d}{d \lambda} \int u dV = 0.
\label{ric-80}
\end{equation}
It means that 
\begin{equation}
	\int u dV = const
\label{ric-90}
\end{equation}
and $u(\lambda)$ represents the probability density of a particle evolving under Brownian motion, backwards in time. It allows us to define the classical, or ``Boltzman - Shannon'' entropy 
\begin{equation}
	S = - \int \limits_{\mathcal M} u \; \mathrm{ln} u \; dV 
\label{ric-100}
\end{equation}
or a renormalized version of the classical entropy 
\begin{equation}
	\tilde{S} = S - \frac{n}{2}  \left \{
		1 + \mathrm{ln} \left[ 4 \pi (\lambda_0 - \lambda) \right] 
	\right \}.
\label{ric-110}
\end{equation}
The $\mathcal W-$functional applied to this backwards Brownian diffusion on a Ricci flow also arises via the renormalized classical entropy $\tilde S$ 
\begin{equation}
	\mathcal W(\lambda) = - \frac{d}{d \lambda} \left( \tau \tilde S \right). 
\label{ric-120}
\end{equation}
We would like to offer the following physical interpretation of the Ricci flow:
\textcolor{blue}{\emph{
\begin{itemize}
	\item the Ricci flow is a statistical system 
	and describes the topology change in quantum gravity
	; 
	\item for every $\lambda$, $g_{ab}(\lambda)$ is a microscopical state in the statistical system; 
	\item $\frac{d}{d \lambda} \mathcal W(g_{ab}, f, \tau)$ is a probability density for the  microscopical state $g_{ab}(\lambda)$.
\end{itemize}
}}
The problem here is that at $\lambda \rightarrow \lambda_0$ (where $\lambda_0$ is some parameter depending on $g_{ab}(\lambda = 0)$) the metric blow up and a singularity appears. The possible solution of this problem is rescaling or renormalizing the Ricci flow. 
\par
The curvature of a Ricci flow blow up in magnitude at a singularity and it is necessary to work towards a theory of ``blowing-up'' that we can rescale a flow more and more as we get closer and closer to a singularity. In Ref.~\cite{anderson} we read: ``The usual method to understand the structure of singularities is to rescale or renormalize the solution on a sequence converging to the singularity to make the solution bounded and try to pass to a limit of the renormalization. Such a limit solution serves as a model for the singularity, and one hopes ... that the singularity models have special features making them much simpler than an arbitrary solution of the equation.'' 
\par 
In Appendix~\ref{app1} the mathematical definitions and theorems are given which are necessary for mathematical understanding of Ricci flow rescaling. For the physical undersctanding of the idea presented here the Theorem~\ref{compactnes_ricci} is necessary only. The theorem states that there exists a regular ``singularity'' model 
($\left( \mathcal N, \hat g_{ab}(t), p \right)$ in the notations of Appendix~\ref{app1}). Such singularity models do in fact have important features making them much simpler than general solutions of the Ricci flow. If $\mathcal N$ is noncompact, then $\mathcal N$ is diffeomophic to $\mathbb R^3, S^2 \times \mathbb R$ or a quotient of these spaces. If $\mathcal N$ is compact, then $\mathcal N$ is diffeomophic to 
$S^3 / \Gamma, S^2 \times S^1$ or $S^2 \times_{\mathbb Z_2} S^1$. 
\par 
Thus we modify the physical interpretation of the Ricci flow as follows:
\textcolor{blue}{\emph{
\begin{itemize}
	\item the Ricci flow is a statistical system 
	and describes the topology change in quantum gravity
	; 
	\item for every $\lambda$, $g_{ab}(\lambda)$ is a microscopical state in the statistical system; 
	\item a probability density for the  microscopical state $g_{ab}(\lambda)$ is defined as $\frac{d}{d \lambda} \hat{\mathcal W}(\hat g_{ab}, f, \tau)$ where 
	$\hat{\mathcal W}(\hat g_{ab}, f, \tau)$ is the Perelman's functional for a rescaled Ricci flow $\left( \mathcal N, \hat g_{ab}(t), p \right)$. 
\end{itemize}
}}
The last item means that the Perelman's functional is calculated on above mentioned singularity model. 

\section{Quantum wormhole in a spacetime foam}
\label{qwhole}

In this section we would like to apply the Ricci flow for the description of appearing/disappearing quantum wormhole in spacetime foam. The strategy of this investigation is following: on the first step we should have a wormhole solution in 4D Einstein gravity with characteristic sizes in Planck region; on the second step we should obtain a Ricci flow with initial conditions as above mentioned  wormhole. 

\subsection{Wormhole supported by two interacting scalar fields}

In Ref.~\cite{Dzhunushaliev:2007cs} it is found a wormhole solution supporting with two phantom scalar fields. The Lagrangian is 
\begin{equation}
  L =-\frac{\mathcal R}{16\pi G} + \epsilon \left[
	  \frac{1}{2}\partial_\mu \varphi \partial^\mu \varphi + 
	  \frac{1}{2}\partial_\mu \chi \partial^\mu
	  \chi-V(\varphi,\chi)
  \right]~,
\label{wh-10}
\end{equation}
where $\mathcal R$ is the 4D scalar curvature, $G$ is the Newtonian gravity constant and the constant $\epsilon = - 1$ means that we consider phantom scalar fields $\phi, \chi$. The potential $V(\varphi,\chi)$ is 
\begin{equation}
	V(\phi,\chi) = \frac{\lambda_1}{4}(\phi^2-m_1^2)^2+
	\frac{\lambda_2}{4}(\chi^2-m_2^2)^2+
	\phi^2\chi^2 - V_0
\label{wh-20}
\end{equation}
where $\phi, \chi$ are two scalar fields with the masses $m_1$ and $m_2$, 
$\lambda_1, \lambda_2$ are the self-coupling constants and $V_0$ - some constant. The field equations are 
\begin{eqnarray}
	\mathcal R_{i}^k - \frac{1}{2} \delta_i^k \mathcal R &=& 8\pi G T^k_i,
\label{wh-30}\\
	\frac{1}{\sqrt{-g}}\frac{\partial}{\partial	x^\mu}\left[
	\sqrt{-g} g^{\mu\nu} \frac{\partial	(\varphi,\chi)}{\partial x^\nu}
	\right] &=& -\frac{\partial V}{\partial	(\varphi,\chi)}.
\label{wh-40}
\end{eqnarray}
where $\mathcal R_{i}^k$ is the 4D Ricci rensor; $T^k_i$ is the energy-momentum tensor for scalar fields $\phi, \chi$; $g_{\mu \nu}$ is 4D spacetime metric \eqref{wh-10}. The wormhole metric is  
\begin{equation}
	ds^2=B(r) dt^2-dr^2-A(r)(d\theta^2+\sin^2\theta d\phi^2)
\label{wh-50}
\end{equation}
where $A(r), B(r)$ are the even functions depending only on the coordinate $r$ which covers the entire range $-\infty < r < +\infty$. Using this metric, one can obtain from Eq's \eqref{wh-30} \eqref{wh-40} the following equations 
\begin{eqnarray}
	\frac{A^{\prime \prime}}{A}-\frac{1}{2}\left(\frac{A^{\prime}}{A}\right)^2-
	\frac{1}{2}\frac{A^{\prime}}{A}\frac{B^{\prime}}{B}&=&
	{\varphi'}^2 + {\chi'}^2 ,
\label{wh-60}\\ 
	\frac{A''}{A}+\frac{1}{2}\frac{A'}{A}\frac{B'}{B} - 
	\frac{1}{2}\left(\frac{A'}{A}\right)^2
	 -\frac{1}{2}\left(\frac{B'}{B}\right)^2+ \frac{B''}{B}&=&
	 2\left[\frac{1}{2}({\varphi'}^2 + {\chi'}^2)+V\right]~, 
\label{wh-70} \\
	\frac{1}{4}\left(\frac{A'}{A}\right)^2-
	\frac{1}{A}+\frac{1}{2}\frac{A'}{A}\frac{B'}{B}&=&
	 -\frac{1}{2}({\varphi'}^2 + {\chi'}^2)+V~,
\label{wh-80}
\end{eqnarray}
where a prime denotes differentiation with respect to $r$. The corresponding field equations from \eqref{wh-40} are 
\begin{eqnarray}
	\varphi''+\left(\frac{A'}{A}+
	\frac{1}{2}\frac{B'}{B}\right)\varphi^\prime &=& 
  \varphi \left[2\chi^2+\lambda_1(\varphi^2-m_1^2)\right]~,  
\label{wh-90}\\
  \chi''+\left(\frac{A'}{A}+\frac{1}{2}\frac{B'}{B}\right)\chi^\prime &=& 
  \chi \left[2\varphi^2+\lambda_2(\chi^2-m_2^2)\right]~.
\label{wh-100}
\end{eqnarray}
In the equations \eqref{wh-60}-\eqref{wh-100} the following rescaling are used: 
$r \rightarrow \sqrt{8\pi G}\, r$, $\varphi \rightarrow \varphi/\sqrt{8\pi G}$,
$\chi \rightarrow \chi/\sqrt{8\pi G}$, $m_{1,2} \rightarrow m_{1,2}/\sqrt{8\pi G}$.
\par 
The boundary conditions are choosing with account of $\mathbb Z_2$ symmetry in the following form:
\begin{alignat}{2}
  \varphi(0)   &=\sqrt{3},        & \qquad \varphi^\prime(0)  &=0, 
\nonumber \\
  \chi(0)      &=\sqrt{0.6},      & \qquad \chi^\prime(0)     &=0, 
\nonumber \\
    A(0)       &=-\frac{1}{V(\phi(0), \chi(0))},              & \qquad A^\prime(0)        &=0, 
\nonumber \\
    B(0)         &=1.0,                    & \qquad B^\prime(0)        &=0,
\label{wh-110}
\end{alignat}
where the condition for $A(0)$ is choosing to satisfy the constraint \eqref{wh-80} at $r=0$, $V(\phi(0), \chi(0))$ is the value of the potential at $r=0$ and the self-coupling  constants $\lambda_1=0.1$ and $\lambda_2=1$.
\par
\begin{figure}[h]
  \includegraphics[width=9cm]{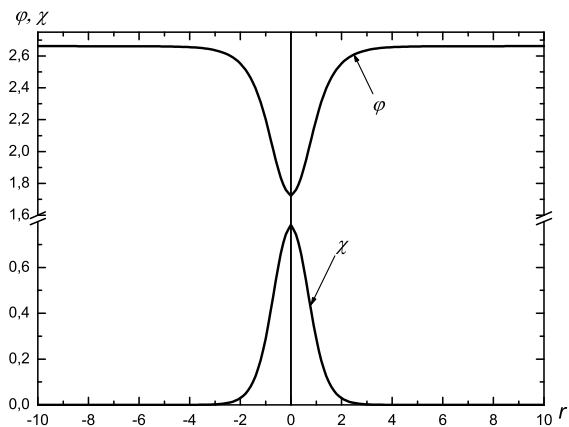}
  \caption{The scalar fields $\varphi, \chi$ in the wormhole model for the boundary conditions given in \eqref{wh-110}.}
\label{phch1}
\end{figure}

\begin{figure}[h]
\begin{minipage}[t]{.49\linewidth}
  \includegraphics[width=9cm]{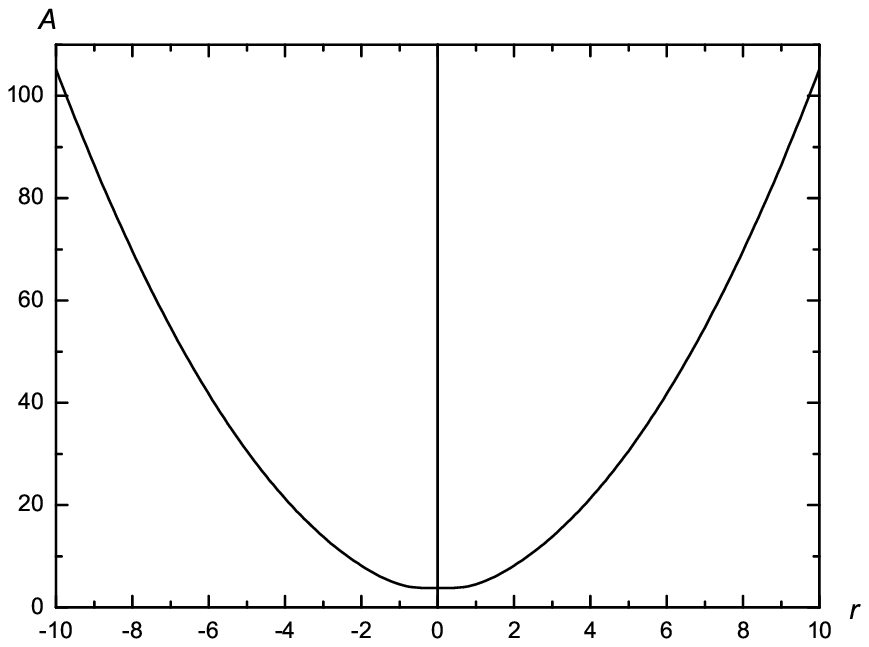}
  \caption{The metric function $A$ in the wormhole model.}
\label{met1}
\end{minipage}\hfill
\begin{minipage}[t]{.49\linewidth}
  \includegraphics[width=9cm]{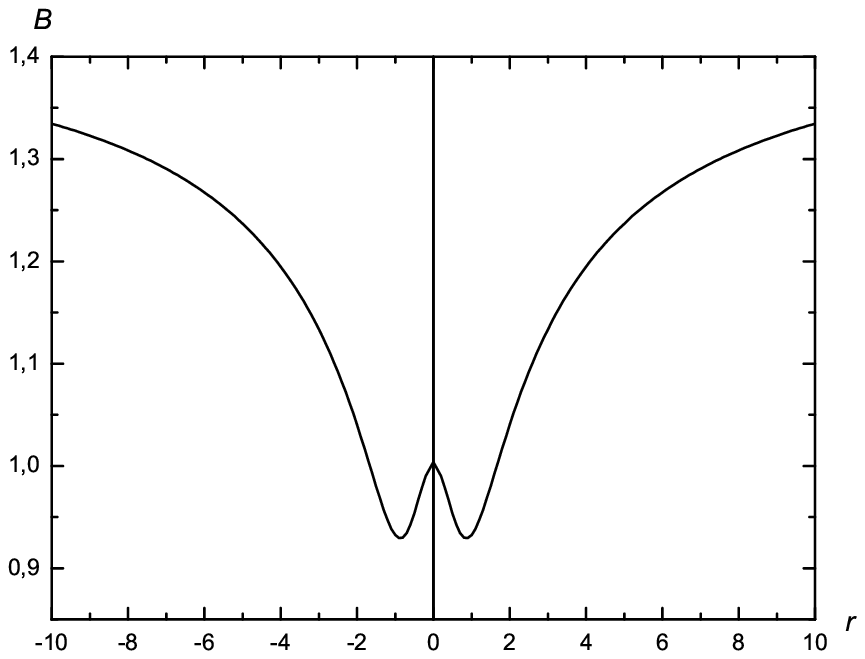}
  \caption{The metric function $B$ in the wormhole model.}
\label{met2}
\end{minipage}\hfill
\end{figure}

 
The results are presented in Fig's.~\ref{phch1}-\ref{met2}. These results are obtained for the masses $m_1\approx 2.661776085$ and $m_2\approx 2.928340304$. 
\par 
The asymptotical behavior of the wormhole solution is 
\begin{eqnarray}
    A &\approx&  \,r^2 + r_0^2, 
\label{wh-112}\\
    B &\approx& B_\infty \left( 1 - \frac{r_0^2}{r^2} \right)
\label{wh-114}
\end{eqnarray}
where $r_0$ and $B_\infty$ are some constants. 

\subsection{Ricci flow started from the wormhole}

We consider 3D part of the 4D wormholw metric \eqref{wh-50} 
\begin{equation}
	dl^2 = e^{2u(r, \lambda)} dr^2 + e^{2v(r, \lambda)} (d\theta^2+\sin^2\theta d\phi^2).
\label{wh-120}
\end{equation}
According to above presented wormhole solution 
\begin{equation}
	u(r, 0) = 0; \; 
	e^{2v(r, 0)} = A(r).
\label{wh-130}
\end{equation}
The corresponding Ricci flow is 
\begin{eqnarray}
	\frac{\partial u}{\partial \lambda} &=& 
	2 e^{-2u} \left(
		v'' - u' v' + {v'}^2
	\right), 
\label{wh-140}\\
	\frac{\partial v}{\partial \lambda} &=& 
		e^{-2u} \left( v'' - u' v' + 2 {v'}^2 \right) - e^{-2 v}. 
\label{wh-150}
\end{eqnarray}

\subsubsection{Ricci soliton}

A Ricci soliton is defined as 
\begin{equation}
	\frac{\partial u}{\partial \lambda} = 
	\frac{\partial v}{\partial \lambda} = 0. 
\label{wh-2-10}
\end{equation}
The solution of Eq's~\eqref{wh-2-10} is 
\begin{equation}
	\left( e^v \right)' = \pm e^u.
\label{wh-2-20}
\end{equation}
In this case one can introduce new coordinate $x$ in following way 
\begin{equation}
	\pm e^u dr = \left( e^v \right)' dr = dx.
\label{wh-2-30}
\end{equation}
After this the metric \eqref{wh-120} is the metric of 3D Euclidean space 
\begin{equation}
	dl^2 = dx^2 + x^2 (d\theta^2+\sin^2\theta d\phi^2).
\label{wh-2-40}
\end{equation}

\subsubsection{Numerical solution}

In this section we would like to present the numerical solution of Eq's~\eqref{wh-140}~\eqref{wh-150}. 

\begin{figure}[h]
\begin{minipage}[t]{.49\linewidth}
  \includegraphics[width=9cm]{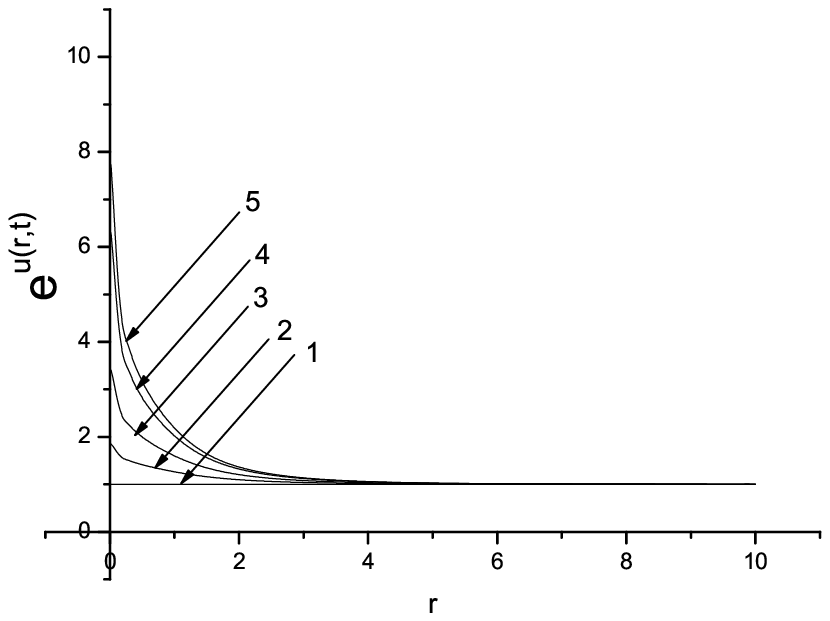}
  \caption{The curves 1,2,3,4,5 ($t_1 = 3.46$) denote correspondingly the profiles 
  $e^{u(r,0)}, e^{u(r, 0.3 t_1)}, e^{u(r, 0.6 t_1)},   e^{u(r, 0.9 t_1)}, e^{u(r, t_1)}$.}
\label{ricci1}
\end{minipage}\hfill
\begin{minipage}[t]{.49\linewidth}
  \includegraphics[width=9cm]{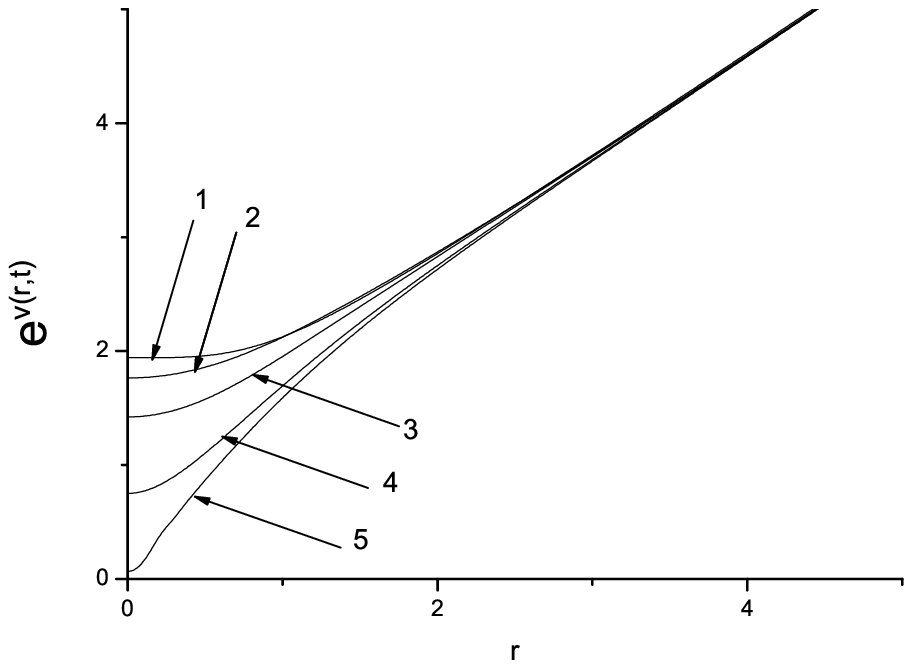}
  \caption{The curves 1,2,3,4,5 ($t_1 = 3.46$) denote correspondingly the profiles 
  $e^{v(r,0)}, e^{v(r, 0.3 t_1)}, e^{v(r, 0.6 t_1)},   e^{v(r, 0.9 t_1)}, e^{v(r, t_1)}$.}
\label{ricci2}
\end{minipage}\hfill
\end{figure}

The boundary conditions are 
\begin{eqnarray}
	u(r, 0) &=& 0; \quad 
	\frac{\partial u(0, \lambda)}{\partial r} = 0; \quad 
	u(\infty , \lambda) = 0; 
\label{wh-160}\\
	v(r, 0) &=& \frac{1}{2} \ln A(r); \quad 
	\frac{\partial v(0, \lambda)}{\partial r} = 0; \quad 
	v(\infty , \lambda) = \ln r. 
\label{wh-170}
\end{eqnarray}
The numerical solution for the Ricci flow is presented in Fig's~\ref{ricci1} and \ref{ricci2}. One can say that this result is in agreement with the investigation presented in Ref.~\cite{Husain:2008rg}: our initial data parameters leads to pinching off of a wormhole mouth. 
We would like to note that in Ref.~\cite{DeBenedictis:2008qm} the topology change is considered as well but not using the Ricci flow .

\section{Discussion and conclusions}

In this paper we have offered the idea about physical interpretation of the Ricci flows. 
The Ricci flow has the statistical interpretation as a quantum wormhole in a spacetime foam.
For every $\lambda$ the metric $g(\lambda)$ is a microscopical state realized with some probability density $\rho(\lambda)$ connected with a Perelman's functional $\hat{\mathcal W}(\lambda)$ of a renormalized Ricci flow. This interpretation is based on the fact that the functional $\hat{\mathcal W}(\lambda)$ is non-decreasing one. Such property allows us to suppose that $\hat{\mathcal W}(\lambda)$ is the probability for the metric $g(\lambda)$ to be in the region $g(\lambda) \in [g(0), g(\lambda)]$. Accordingly 
$\frac{d \hat{\mathcal W}(\lambda)}{d \lambda}$ is proportional to corresponding probability density. 
\par
We would like to list the problems for the future investigation in this direction:
\begin{enumerate}
	\item The Ricci flow considered in section~\ref{qwhole} is not covariant from 4D point of view. Consequently it is necessary to investigate the question about 4D covariance of the Ricci flow.
	\item It is necessary to calculate a Perelman's functional for a rescaled Ricci flow. 
	\item The Ricci flows considered in Ref.~\cite{topping} are defined on compact manifolds but the quantum wormholes are non-compact manifolds. Consequently it is necessary to have in non-compact case the theorem similar to compactness theorem \ref{compactnes_ricci} of Ricci flows.
	\item In the statistical mechanics the probability density of a microscopical state is connected with Hamiltionian of some physical system. The question is: there exists some physical system whose statistical properies leads to the Perelman's functional ?
\end{enumerate}
For the item (1) we may note that probably the rescaled Ricci flow does not depend on 3+1 decomposition which we have used in section~\ref{qwhole}. It is possible that in this case we will have 4D invariance. 
\par 
In conclusion we would like to emphasize that we consider only the problem connected with the topology change in quantum gravity. In our opinion this problem is not connected with the problem of non-perturbative field quantization of a gravitational field (metric).

\section{Acknowledgments}
I am grateful to V. Vanchurin for the fruitful discussion about the statistical interpretation of the Ricci flow.

\appendix
\section{Rescaling of a Ricci flow}
\label{app1}

In this section we follow to Ref.\cite{topping}. 
\par 
\begin{definition}
A sequence $\left( \mathcal M_i, \left( g_{ab} \right)_i, p_i \right)$ of smooth, complete, \emph{pointed} Riemannian manifolds (that is, Riemannian manifolds $\left( \mathcal M_i, \left( g_{ab} \right)_i \right)$ and points $p_i \in \mathcal M$) is said to \emph{converge} (smoothly) to the smooth, complete, pointed manifold $\left( \mathcal M, g_{ab}, p \right)$ as $i \rightarrow \infty$ if there exist 
\begin{enumerate}
	\item a sequence of compact sets $\Omega_i \in \mathcal M$, exhausting $\mathcal M$ (that is, so that any compact set $K \subset \mathcal M$ satisfies $K \subset \Omega_i$ for sufficiently large $i$) with $p \in \mathrm{int} \left( \Omega_i \right)$ for each $i$;
	\item a sequence of smooth maps $\phi_i : \Omega_i \rightarrow \mathcal M_i$ which are diffeomerphic onto their image and satisfy $\phi_i(p) = p_i$ for all $i$; 
\end{enumerate}
such that, 
\begin{equation}
	\phi_i^* \left( g_{ab} \right)_i \rightarrow g_{ab}
\label{app-10}
\end{equation}
smoothly as $i \rightarrow \infty$ in the sense that for all compact sets $K \subset \mathcal M$, the tensor $\phi_i^* \left( g_{ab} \right)_i - g_{ab}$ and its covariant derivatives of all orders (which respect to any fixed background connection) each converge uniformly to zero on $K$.
\end{definition}
\par 
Two consequences of the convergence 
$\left( \mathcal M_i, \left( g_{ab} \right)_i, p_i \right) \rightarrow 
\left( \mathcal M, g_{ab}, p \right)$ are that 
\begin{enumerate}
	\item for all $s > 0$ and $k \in \{ 0 \} \cup \mathbb N$, 
\begin{equation}
	\sup \limits_{i \in \mathbb N} 
	\sup \limits_{B_{g_i}(p_i, s)} \left|
		\nabla^a R^b_{cde}\left( g_i \right)
	\right| < \infty ;
\label{app-20}
\end{equation}
	\item 
\begin{equation}
	\inf \limits_i \; \mathrm{inj} 
	\left( \mathcal M_i, \left( g_{ab} \right)_i, p_i \right) > 0 , 
\label{app-30}
\end{equation}
where $\mathrm{inj} \left( \mathcal M_i, \left( g_{ab} \right)_i, p_i \right)$ denotes the injectivity radius of $\left( \mathcal M_i, \left( g_{ab} \right)_i\right)$ at $p_i$.
\end{enumerate}
\begin{theorem}
(Compactness - manifolds). 
Suppose that $\left( \mathcal M_i, \left( g_{ab} \right)_i, p_i \right)$ is a sequence of complete, smooth, pointed Riemannian manifolds (all of dimension $n$) satisfying \eqref{app-20} and \eqref{app-30}. Then exists a complete, smooth, pointed Riemannian manifold 
$\left( \mathcal M, g_{ab}, p \right)$ (of dimension $n$) such that after passing to some subsequence in $i$, 
\begin{equation}
	\left( \mathcal M_i, \left( g_{ab} \right)_i, p_i \right) \rightarrow 
	\left( \mathcal M, g_{ab}, p \right)
\label{app-40}
\end{equation}
\end{theorem}
One can derive, from the compactness theorem for manifolds (theorem 1) a compactness theorem for Ricci flows. 
\begin{theorem}
Let $\left( \mathcal M_i, \left( g_{ab} \right)_i(t) \right)$ be a sequence of smooth families of complete Riemannian manifolds for $t \in (a,b)$ where 
$-\infty \leq a < 0 < b \leq \infty$. Let $p_i \in \mathcal M_i$ for each $i$. Let 
$\left( \mathcal M, g_{ab}(t) \right)$ be a smooth family of complete Riemannian manifolds for $t \in (a,b)$ and let $p \in \mathcal M$. We say that 
\begin{equation}
	\left( \mathcal M_i, \left( g_{ab} \right)_i(t), p_i \right) \rightarrow 
	\left( \mathcal M, g_{ab}(t), p \right)
\label{app-50}
\end{equation}
as $i \rightarrow \infty$ if there exist 
\begin{enumerate}
	\item a sequence of compact $\Omega_i \subset \mathcal M$ exhausting $\mathcal M$ and satisfying $p \in \mathrm{int} \left( \Omega_i \right)$ for each $i$; 
	\item a sequence of smooth maps $\phi_i : \Omega_i \rightarrow \mathcal M_i$, diffeomorphic onto their image, and with $\phi(p_i) = p_i$;
\end{enumerate}
such that 
\begin{equation}
	\phi_i^* \left( g_{ab} \right)_i(t) \rightarrow g_{ab}(t) 
\label{app-60}
\end{equation}
as $i \rightarrow \infty$ in the sense that 
$\phi_i^* \left( g_{ab} \right)_i(t) - g_{ab}(t)$ and its derivatives of every order (with respect to time as well as covariant space derivatives with respect to any fixed background connection) converge uniformly to zero on every compact subset of $\mathcal M \times (a,b)$. 
\end{theorem}
One can prove the following result 
\begin{theorem}
\label{compactnes_ricci}
(Compactness of Ricci flows.) 
Let $\mathcal M_i$ be a sequence of manifolds of dimension $n$, and let $p_i \in \mathcal M_i$ for each $i$. Suppose that $\left( g_{ab} \right)_i$ is a sequence of complete Ricci flows on 
$\mathcal M_i$ for $t \in (a,b)$, where $-\infty \leq a < 0 < b \leq \infty$. Suppose that 
\begin{enumerate}
	\item 
\begin{equation}
	\sup \limits_{i} \; 
	\sup \limits_{x \in \mathcal M_i, t \in (a,b)} \left|
		R^a_{bcd}\left( g_i(t) \right)
	\right|(x) < \infty ;
\label{app-70}
\end{equation}
	\item 
\begin{equation}
	\inf \limits_i \; \mathrm{inj} 
	\left( \mathcal M_i, \left( g_{ab} \right)_i(0), p_i \right) > 0 .
\label{app-80}
\end{equation}
\end{enumerate}
Then there exist a manifold $\mathcal M$ of dimension $n$, a complete Ricci flow $g_{ab}(\lambda)$ on $\mathcal M$ for $t \in (a,b)$, and a point $p \in \mathcal M$ such that, after passing to a subsequence in $i$, 
\begin{equation}
	\left( \mathcal M_i, \left( g_{ab} \right)_i(t), p_i \right) \rightarrow 
	\left( \mathcal M, g_{ab}(t), p \right)
\label{app-90}
\end{equation}
as $i \rightarrow \infty$. 
\end{theorem}
The application of the compactness of Ricci flows in Theorem~\ref{compactnes_ricci} is to analyze rescaling of Ricci flows near their singularities. Let 
$\left( \mathcal M, g_{ab}(t) \right)$ be a Ricci flow with $\mathcal M$ closed, on the maximal interval $[0, \lambda_0)$. In the consequence of a singularity 
\begin{equation}
	\sup \limits_{\mathcal M} \left| R^a_{bcd} \right|(\cdot, \lambda) \rightarrow \infty 
\label{app-100}
\end{equation}
as $\lambda \rightarrow \lambda_0$. Let us choose points $p_i \in \mathcal M$ and 
$\lambda_i \rightarrow \lambda_0$ such that 
\begin{equation}
	\left| R^a_{bcd} \right|(p_i, \lambda_i) = 
	\sup \limits_{x \in \mathcal M, \lambda \in [0, \lambda_i]} 
	\left| R^a_{bcd} \right|(x, \lambda).
\label{app-110}
\end{equation}
Define rescaled (and translated) flows $(g_{ab})_i(\lambda)$ by 
\begin{equation}
	(g_{ab})_i(\lambda) = \left| R^a_{bcd} \right|(p_i, \lambda_i) 
	g_{ab} \left[
		\lambda_i + \frac{\lambda}{\left| R^a_{bcd} \right|(p_i, \lambda_i)}
	\right]
\label{app-120}
\end{equation}
One can show that $\left( \mathcal M, (g_{ab})_i \right)$ is a Ricci flow on the interval 
$\biggl[ -\lambda_i \left| R^a_{bcd} \right|(p_i, \lambda_i), 
\left( \lambda_0 - \lambda_i \right)\left| R^a_{bcd} \right|(p_i, \lambda_i) \biggl]$. 
\par 
One can show that for all $a < 0$ and some $b > 0$, $(g_{ab})_i)$ is defined for 
$\lambda \in (a,b)$ and 
\begin{equation}
	\sup \limits_i \sup \limits_{\mathcal M \times (a,b)} 
	\left| R^a_{bcd} \left( (g_{ab})_i \right) \right| < \infty. 
\label{app-130}
\end{equation}
By Theorem~\ref{compactnes_ricci} one can pass to a subsequence in $i$, and get convergence 
$\left( \mathcal M, (g_{ab})_i(\lambda), p_i \right) \rightarrow 
\left( \mathcal N, \hat g_{ab}(\lambda), p_\infty \right)$ to a ``singularity model'' Ricci flow $\left( \mathcal N, \hat g_{ab}(\lambda) \right)$, provided that we can establish the injectivity radius estimate 
\begin{equation}
	\inf \limits_i \mathrm{inj} \left(
		\mathcal M, (g_{ab})_i(0), p_i)
	\right) > 0. 
\label{app-140}
\end{equation}

\end{document}